%% file: paper_MRI_prl_arxiv.tex
\documentclass[amsmath,amssymb,prl,twocolumn,superscriptaddress,nofootinbib]{revtex4-2}

\usepackage{graphicx,natbib}
\usepackage{dcolumn}
\usepackage{bm}
\usepackage{hyperref}
\usepackage{xcolor}

\newcommand\bb[1]{\mbox{\boldmath{$#1$}}}
\newcommand\grad{\bb{\nabla}}
\newcommand\bcdot{\,\bb{\cdot}\,}
\newcommand\bdbldot{\,\bb{:}\,}
\newcommand\btimes{\,\bb{\times}\,}

\newcommand{\rmd}{\textrm{d}}

\newcommand{\vA}{v_{\mathrm{A},0}}
\newcommand{\Om}{\Omega_0}

\newcommand{\omCz}{\omega_{\mathrm{c},0}}
\newcommand{\omp}{\omega_\mathrm{p}}
\newcommand{\omrat}{\omega_{\mathrm{c},0}/\Omega_0}
\newcommand{\rhoC}{\rho_\mathrm{c}}
\newcommand{\rhoCz}{\rho_{\mathrm{c},0}}
\newcommand{\lmri}{\lambda_\mathrm{MRI}}
\newcommand{\vecv}{\bb{v}} 
\newcommand{\vecE}{\bb{E}} 
\newcommand{\vecB}{\bb{B}} 
\newcommand{\vecu}{\bb{u}}
\newcommand{\vecV}{\bb{V}}
\newcommand{\vecx}{\bb{x}}

\newcommand{\ppar}{p_{\|}}
\newcommand{\pperp}{p_\perp}
\DeclareMathAlphabet\mathbfcal{OMS}{cmsy}{b}{n}
\hypersetup{
    colorlinks,
    linkcolor={red!50!black},
    citecolor={blue!50!black},
    urlcolor={blue!80!black}
}
\colorlet{RED}{red}

\newcommand{\prlsection}[1]{\emph{#1.}---}

\begin{document}

\title{Collisionless Magnetorotational Turbulence in Pair Plasmas: \\
Steady-state Dynamics, Particle Acceleration, and Radiative Cooling}

\author{Fabio Bacchini}
\email{fabio.bacchini@kuleuven.be}
\affiliation{Centre for Mathematical Plasma Astrophysics, Department of Mathematics, KU Leuven, Celestijnenlaan 200B, B-3001 Leuven, Belgium}
\affiliation{Royal Belgian Institute for Space Aeronomy, Solar-Terrestrial Centre of Excellence, Ringlaan 3, 1180 Uccle, Belgium}
\author{Vladimir Zhdankin}
\affiliation{Department of Physics, University of Wisconsin-Madison, Madison, WI 53706, USA}
\affiliation{Center for Computational Astrophysics, Flatiron Institute, 162 Fifth Avenue, New York, NY 10010, USA}
\author{Evgeny A.\ Gorbunov}
\affiliation{Centre for mathematical Plasma Astrophysics, Department of Mathematics, KU Leuven, Celestijnenlaan 200B, B-3001 Leuven, Belgium}
\author{Gregory R.\ Werner}
\affiliation{Center for Integrated Plasma Studies, Department of Physics, University of Colorado, 390 UCB, Boulder, CO 80309-0390, USA}
\author{Lev Arzamasskiy}
\affiliation{School of Natural Sciences, Institute for Advanced Study, Princeton, NJ 08544, USA}
\author{Mitchell C.\ Begelman}
\affiliation{JILA, University of Colorado and National Institute of Standards and Technology, 440 UCB, Boulder, CO 80309-0440, USA}
\affiliation{Department of Astrophysical and Planetary Sciences, University of Colorado, 391 UCB, Boulder, CO 80309-0391, USA}
\author{Dmitri A.\ Uzdensky}
\affiliation{Center for Integrated Plasma Studies, Department of Physics, University of Colorado, 390 UCB, Boulder, CO 80309-0390, USA}

\begin{abstract}
We present 3D fully kinetic shearing-box simulations of pair-plasma magnetorotational turbulence with unprecedented macro-to-microscopic scale separation.
While retrieving the expected fluid behavior of the plasma at large scales, we observe a steepening of turbulent spectra at kinetic scales and substantial angular-momentum transport linked with kinetic processes.
For the first time, we provide a definitive demonstration of nonthermal particle acceleration in kinetic magnetorotational turbulence agnostically of initial conditions, by means of a novel strategy exploiting synchrotron cooling.
\end{abstract}


\maketitle

\prlsection{Introduction}
Recent years have witnessed a surge of interest in the plasma and astrophysics communities toward high-energy environments such as the surroundings of supermassive black holes (SMBHs). Groundbreaking observations with the Event Horizon Telescope \cite{EHT2019e,EHT2022e} and Gravity \cite{gravity2019} hinted at the existence of (quasi)stationary plasma structures (e.g.\ disks) around SMBHs, underlining the importance of modeling plasma flows in strong-gravity environments. Most efforts in this direction have relied on magnetohydrodynamic (MHD) simulations \cite{gammie2003,narayan2012,porth2019,tomei2019,dexter2020,ripperda2020,ripperda2022}, often resorting to relatively simple prescriptions to include limited microscopic (i.e.\ kinetic) effects \cite{ressler2015,chael2018,dexter2021,hankla2022,scepi2022}. However, it is commonly accepted that plasmas around low-luminosity SMBHs are collisionless, and that kinetic plasma physics can shape the global dynamics, thereby invalidating basic MHD assumptions. For example, accretion onto low-luminosity SMBHs is subject to enhanced angular momentum transport (AMT) by beyond-MHD collisionless turbulence \cite{sharma2006,kempski2019}. Moreover, electron-ion plasmas in disks likely exist in two-temperature states where relativistic electrons produce most of the radiation (e.g.\ via synchrotron cooling, SC). Finally, relativistic electron energies could result from nonthermal particle acceleration (NTPA) via turbulent dissipation of magnetic fields. MHD models cannot completely capture these effects, rendering fully kinetic simulations of accreting plasmas necessary.

Due to computational constraints, only \textit{local} Particle-in-Cell (PIC) kinetic simulations of turbulent accretion (e.g.\ considering individual disk-plasma patches) are possible to date \cite{riquelme2012,hoshino2013,hoshino2015,inchingolo2018,bacchini2022mri} (with recent, important progress for large-scale/global modeling \cite{parfrey2019,crinquand2020,elmellah2022,galishnikova2023,sandoval2023}). The shearing-box (SB) paradigm is a reliable tool for such local simulations \cite{hill1878,hawley1995,matsumototajima1995,stone1996,sanoinutsuka2001,baistone2013,hirai2018} focusing on the magnetorotational instability (MRI)  \cite{balbushawley1991,goodmanxu1994,pessahgoodman2009,goedbloedkeppens2022}, whose nonlinear development drives turbulence. However, the handful of existing PIC-SB MRI simulations suffer from severe limitations \cite{riquelme2012,hoshino2013,hoshino2015,inchingolo2018,bacchini2022mri}. Of particular importance is scale ordering: for realism, macroscopic (MRI--MHD) scales should be well separated from kinetic plasma scales, but PIC simulations often employ rescaled parameters to limit computing costs. This can dramatically alter the MRI development and produce large deviations in overall dynamics between MHD and PIC-SB runs. Recently, we demonstrated that ``large-enough'' 3D PIC simulations retrieve the expected MHD-like dynamics \cite{bacchini2022mri}, attaining a saturated, nonlinear turbulent state. However, it is still essential to explore larger scale separations to definitively determine the MHD-to-kinetic transition of MRI turbulence in magnetic and bulk-velocity power spectra. Limited scale separation in earlier 3D works also caused the unrealistically slow growth of kinetic (especially mirror) instabilities regulating pressure anisotropy, thereby affecting overall AMT.

Importantly, previous PIC-SB MRI simulations \cite{riquelme2012,hoshino2013,hoshino2015,inchingolo2018,bacchini2022mri} did not distinguish NTPA in the initial transient stages (including linear MRI growth) that precede fully developed MRI turbulence. These ``preturbulent'' stages develop macroscopic (system-scale) current sheets that undergo magnetic reconnection, releasing large amounts of energy and producing considerable NTPA in PIC simulations; however, these early stages are artificial (i.e.\ only related to the development of the MRI in the SB paradigm) and hence irrelevant for realistic accretion disks, and should therefore be excluded. Indeed, we assert that the capability of MRI turbulence to accelerate particles \emph{has not been assessed at all thus far} in a manner that clearly distinguishes turbulent energization from earlier artificial injection mechanisms.

In this Letter, we advance the frontier of collisionless-MRI studies by performing the largest (to date) 3D PIC-SB simulations. We employ unprecedented scale separations to obtain PIC results retrieving expected MHD dynamics, and allowing us to study detailed microphysical effects potentially affecting the global accretion flow. Moreover, by imposing radiative cooling to suppress NTPA during (only) the preturbulence stages, we distinctly characterize NTPA \textit{solely due to MRI turbulence}. We also show that cooling (if kept active only during the preturbulence stages) has negligible effect on the MRI evolution aside from extinguishing the high-energy nonthermal particles; this is the first exploration of radiation-reaction effects in PIC-SB simulations.

\prlsection{Numerical Model and Setup} We conduct 3D PIC-SB simulations with \textsc{Zeltron} \cite{cerutti2013}, employing the kinetic SB with orbital advection (KSB-OA) \cite{bacchini2022mri} to model a local Cartesian ($x\equiv\,$radial direction from the central SMBH, $y\equiv\,$toroidal, $z\equiv\,$vertical) sector of an accretion disk\footnote{Since the publication of the original KSB-OA method \cite{bacchini2022mri}, we have improved the SB implementation in \textsc{Zeltron} with a more stable and less diffusive treatment of the shearing-periodic boundary conditions.}. The local plasma patch represented by the simulation box has orbital frequency $\Om$ and is subjected to a background force corresponding to a (linearized) Keplerian shearing-velocity profile $\vecv_\mathrm{s}(x)=-(3/2)\Om x\bb{\hat{y}}$. We apply shearing-periodic boundary conditions at $x$-boundaries, while $y$- and $z$-boundaries are purely periodic~\cite{bacchini2022mri}. Our initial setup consists of a thermal pair plasma for simplicity ($m=m_i=m_e$), with uniform per-species number density $n_0=n_{i,0}=n_{e,0}$, nonrelativistic uniform (normalized) temperature $\theta_0\equiv k_\mathrm{B}T_0/(mc^2) \ll 1$ (with $T_0=T_{i,0}=T_{e,0}$), and zero (fluctuating) bulk speed. A weak, purely vertical magnetic field $\vecB_0=(0,0,B_0)$ initially threads the plasma. The free parameters of our model are therefore the physical box size $L_x\times L_y\times L_z$, the initial temperature $\theta_0$, the initial Alfv\'en-to-light speed ratio (including both particle species) $\vA/c\equiv B_0/\sqrt{8\pi n_0 mc^2}$, and the initial cyclotron-to-rotational frequency ratio $\omrat$ where $\omCz\equiv eB_0/(mc)$. On the basis of previous work \cite{bacchini2022mri}, we choose $L_z=2\lmri$ (with $\lmri\equiv 2\pi\vA/\Om$ the most-unstable MRI wavelength), $\theta_0=1/128$, $\vA/c\simeq0.007$, and $\omrat = 30$ to obtain the nonrelativistic length-scale ordering\footnote{In realistic systems, it is expected that $H_0/\rhoC\propto\omega_\mathrm{c}/\Omega_0$ can reach extreme values, e.g.\ $10^7\mbox{--}10^9$ around M87*.} $c/\omp<\rhoCz<\lmri<L_z\le H_0$; here, $\omp\equiv\sqrt{8\pi e^2n_0/m}$ is the total plasma frequency, $\rhoCz\equiv\sqrt{\theta_0} mc^2/(eB_0)$ is the (initial) typical particle gyroradius, and $H_0\equiv\sqrt{\theta_0}c/\Om$ is the disk scale height. These parameters thus determine a separation $\rhoCz/(c/\omp)\simeq 12.5$, $\lmri/\rhoCz\simeq15$, and $H_0/\lmri=L_z/\lmri=2$ between the macroscopic and microscopic length scales involved. Our choice of parameters results in an initial plasma $\beta_0\equiv2n_0k_\mathrm{B} T_0/[B_0^2/(8\pi)] =2\theta_0/(\vA/c)^2 = 2(\rho_{\mathrm{c},0}/(c/\omp))^2\simeq312$. Additionally, we choose $L_x=2L_z$ such that the shearing-velocity offset between $x$-boundaries ($\pm L_x/2$) is $(3/2)\Om L_x<c$, and we set $L_y/L_z=4$. We employ 54 particles per cell with grid spacing $\Delta x\simeq (c/\omp)/2$, and 8 current-filtering passes per time step, to combat numerical noise.

To model SC in MRI turbulence, we add a synchrotron-cooling force to the right-hand side of the KSB-OA particle equations of motion \cite{bacchini2022mri}. As in several other works including SC in PIC \cite{tamburini2011,cerutti2013,cerutti2016,comissosironi2021,hakobyan2023apjl,schoeffler2023}, we write this term as
\begin{equation}
    \left(\frac{\rmd\vecu}{\rmd t}\right)_\mathrm{SC} = -\kappa\gamma\left[\left(\vecE+\frac{\vecu}{\gamma}\btimes\vecB\right)^2-\left(\frac{\vecu}{\gamma}\bcdot\vecE\right)^2\right]\vecu,
\end{equation}
where $\vecu$ is the spatial part of the particle four-velocity, $\gamma\equiv\sqrt{1+u^2/c^2}$ is the particle Lorentz factor, $\vecE$ is the electric field, and $\kappa$ is tuned such that SC maintains a quasithermal state at $\gamma-1\sim 1$. The MRI in our setup grows out of numerical noise; we evolve the initial state for a total time $t_\mathrm{end}=15P_0$ ($P_0\equiv 2\pi/\Om$).

\begin{figure*}
\centering
\includegraphics[width=1\textwidth, trim={0mm 0mm 0mm 0mm}, clip]{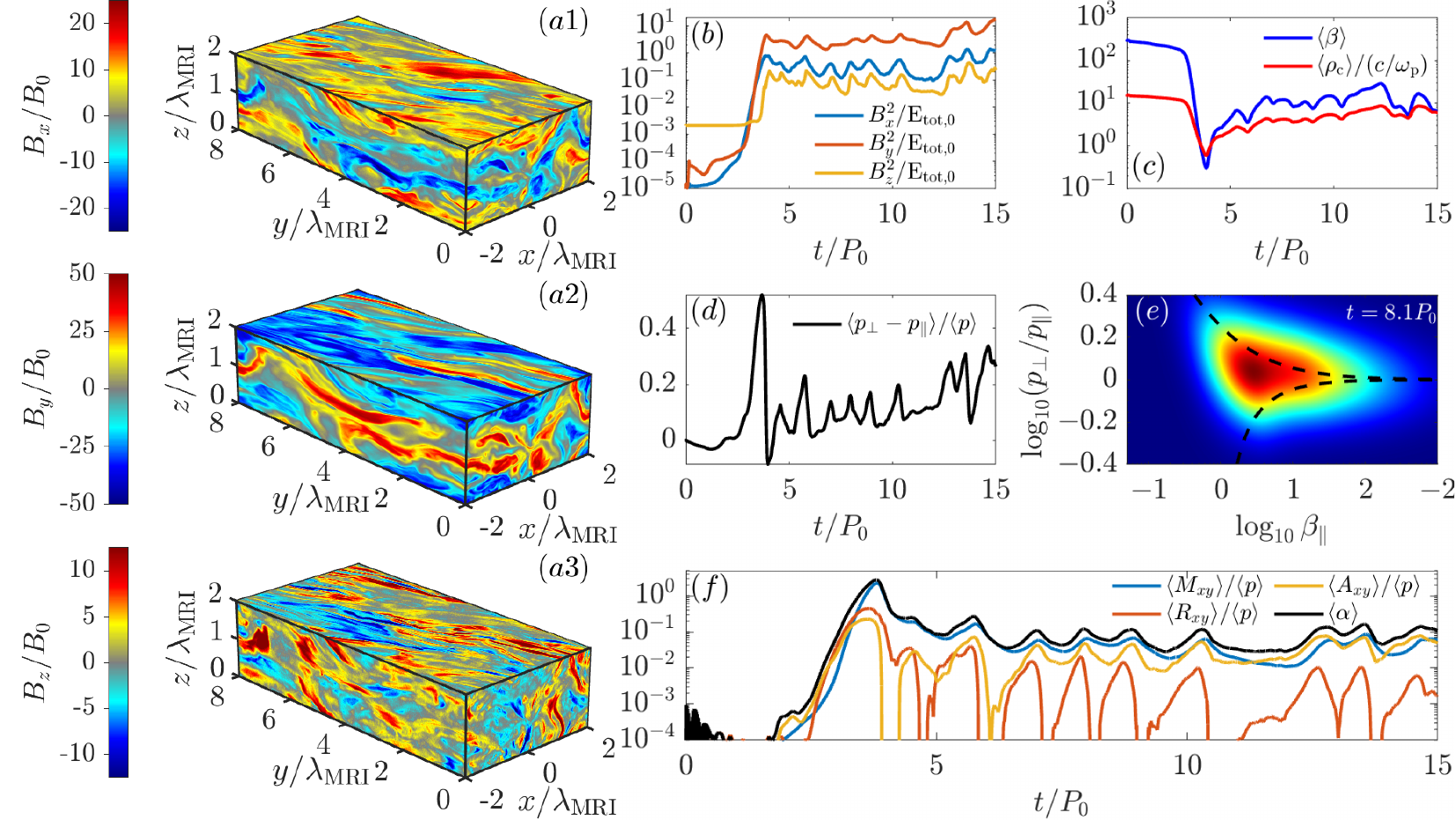}
\caption{Properties of mesoscale MRI turbulence. $(a1\mbox{--}3)$: Component-wise magnetic-field distribution at $t=8.1P_0$; $(b)\mbox{--}(c)$: Average magnetic-field amplification, plasma-$\beta$, and gyroradius throughout the run; $(d)\mbox{--}(e)$: Average pressure anisotropy over time and $(\beta_\|,p_\perp/p_\|)$-distribution with respect to mirror/firehose approximate thresholds during the nonlinear stage; $(f)$: $\alpha$-model stresses (contributing to AMT) over time.}
\label{fig:fields_energy}
\end{figure*}

\begin{figure*}
\centering
\includegraphics[width=1\textwidth, trim={0mm 0mm 0mm 0mm}, clip]{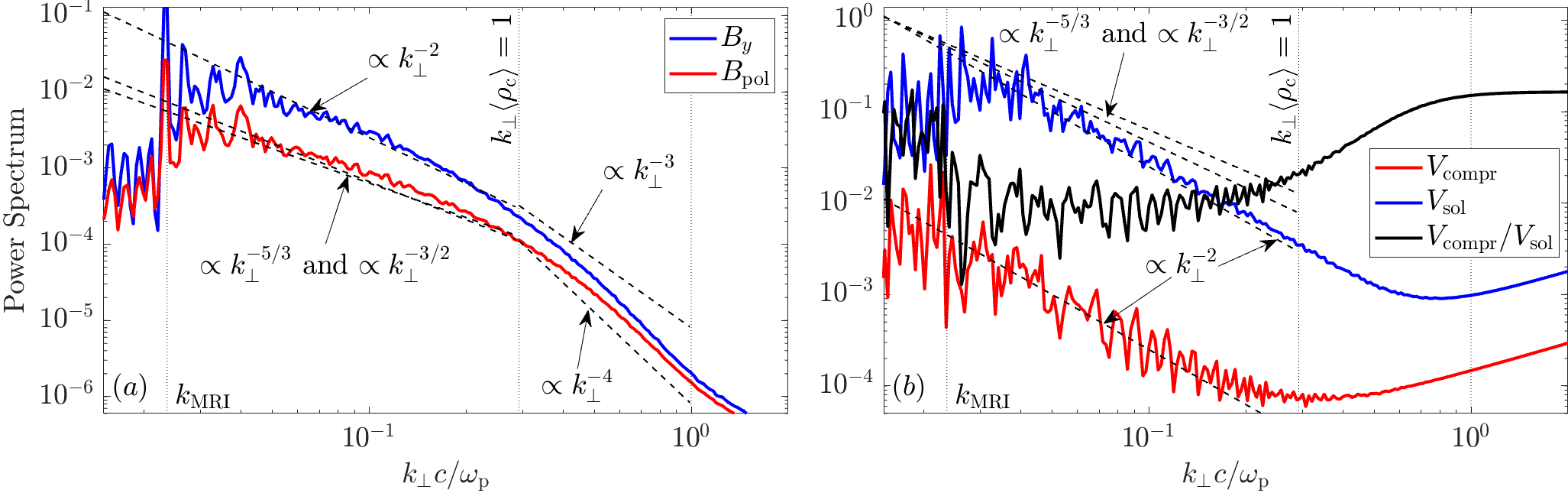}
\caption{$(a)$: Toroidal and poloidal magnetic-field power spectra versus $k_\perp$ averaged over $t\in[7,8]P_0$ during the nonlinear stage; $(b)$: Likewise for compressive and solenoidal velocity-fluctuation spectra.}
\label{fig:fft}
\end{figure*}

\begin{figure}
\centering
\includegraphics[width=1\columnwidth, trim={0mm 0mm 0mm 0mm}, clip]{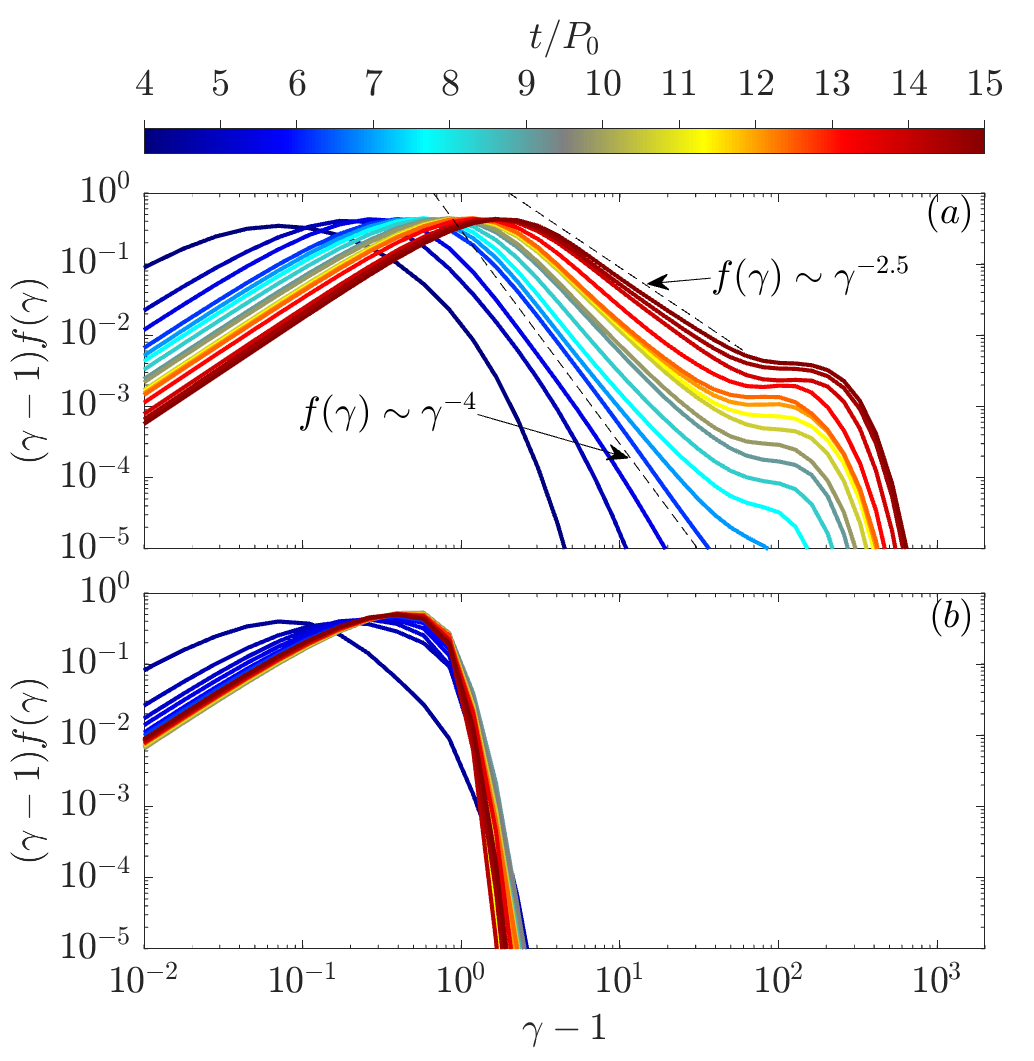}
\caption{Particle energy distributions throughout the MRI nonlinear stage, $(a)$ without SC and $(b)$ with SC activated for the whole run.}
\label{fig:dist}
\end{figure}

\prlsection{Results}
Via background shearing and small-scale perturbations, initially vertical magnetic-field lines bend and $\vecB$ acquires $x,y$-components. The system's evolution encompasses an initial stage, where small-scale perturbations are excited; an exponential magnetic-field growth due to linear MRI modes; and a nonlinear, saturated stage where primary-MRI channel flows decay into a turbulent state. The channel decay is driven by parasitic instabilities (e.g.\ tearing and kinking of current sheets at channel interfaces) feeding off the primary MRI \cite{goodmanxu1994,pessahgoodman2009}. These instabilities require sufficiently large domains to develop, and can be suppressed if specific conditions are not met~\cite{latter2009}; our system size ensures that current-sheet kinking and tearing efficiently destroy the primary MRI channels and initiate the nonlinearly turbulent stage~\cite{bacchini2022mri}. A typical spatial distribution of magnetic-field components during the nonlinear stage is shown in Fig.~\ref{fig:fields_energy} (panels $(a1\mbox{--}3)$), where we observe the chaotic configuration of the MRI-driven turbulence. As outlined above, we inhibit NTPA until the turbulent state is established by keeping SC active until right after the initial channel disruption\footnote{Smaller simulations show that deactivating SC at later times produces qualitatively similar results.}. At $t\simeq4P_0$, we remove SC, such that particles can be energized by the MRI turbulence only, starting from an approximately Maxwellian state (see below).

The right panels in Fig.~\ref{fig:fields_energy} show the evolution of the physical properties of MRI turbulence. Panel $(b)$ shows the evolution of each component of magnetic-field energy, averaged over space. The exponential amplification of magnetic fields, due to the primary MRI, can be observed until $t\sim4P_0$. At that time, the MRI channel flows decay into a sustained-turbulence state with a statistically constant total magnetic energy. This quasi-steady state is due to a balance between constant energy injection by the MRI and turbulent energy cascade. When the turbulence reaches kinetic scales, plasma heating and high-energy particle acceleration can occur. Panel $(c)$ shows that the volume-averaged $\beta$ and average particle gyroradius slowly increase during the nonlinear stage, as a consequence of turbulent heating\footnote{Note that $\beta\sim10$ during the whole nonlinear stage, implying that turbulent models of disk plasmas should consider a high-$\beta$ collisionless regime~\cite{arzamasskiy2023}.}. In this turbulent state, we can evaluate AMT throughout the disk, e.g.,\ via the $\alpha$-parametrization \cite{shakurasunyaev1973}, involving the Maxwell ($M_{xy}\equiv-B_xB_y/(4\pi)$), Reynolds ($R_{xy}$), and anisotropic ($A_{xy}$) stresses. Here, $R_{xy} \equiv mn U_x U_y/\Gamma^2$ is subdominant due to the lack of macroscopically coherent flows in the turbulent accretion disk (with density $n$ and bulk velocity $\bb{U}\equiv(1/n)\int \vecu f(\vecx,\vecu,t) \rmd^3 \vecu$ calculated from the distribution function, $f$, and $\Gamma\equiv\sqrt{1+U^2/c^2}$). We evaluate the anisotropic stress $A_{xy} \equiv -(p_\perp-p_\|)B_xB_y/B^2$ \cite{kulsrud1983} by computing parallel and perpendicular plasma pressure, $\ppar\equiv\textbf{P}\bb\bdbldot\vecB\vecB/B^2$ and $\pperp\equiv\textbf{P}\bdbldot(\mathbb{I}-\vecB\vecB/B^2)/2$, where $\textbf{P} \equiv m\int (\vecu\vecu/\gamma)f(\vecx,\vecu,t) \rmd^3 \vecu$ is the full pressure tensor; $A_{xy}$ does not exist in standard MHD calculations with isotropic pressure. In panel~$(d)$, we observe a positive volume-averaged pressure anisotropy $\langle\pperp-\ppar\rangle>0$ during the entire nonlinear stage, indicating the presence of perpendicular-pressure enhancement mechanisms (e.g., approximate conservation of the first adiabatic invariant $p_\perp/(mnB)$ in growing fields). Perpendicular-pressure growth is limited by kinetic effects, as shown in panel $(e)$ by the $(\beta_\|,p_\perp/p_\|)$-distribution (where $\beta_\|\equiv p_\|/(B^2/(8\pi))$) at a representative time $t=8.1P_0$ during the turbulent stage. Here, we observe the presence of mirror modes: as $\pperp$ increases, approximate thresholds for mirror instabilities (top dashed line) \cite{kunz2014} are crossed, scattering particles efficiently and limiting further increase in anisotropy. Similarly, the firehose instability (bottom dashed line) \cite{kunz2014} prevents the growth of $\ppar$ (due to turbulent reconnection and/or conservation of the adiabatic invariant $p_\| B^2/(mn)^3$ \cite{hoshino2015} when $B$ is dissipated). Panel $(f)$ shows our first-principles measurement of AMT: we observe that pressure anisotropy does indeed substantially contribute to~$\alpha$, via $\langle A_{xy}\rangle\sim\langle M_{xy}\rangle$ at all times. The resulting $\langle\alpha\rangle\equiv\langle R_{xy}+A_{xy}+M_{xy}\rangle/\langle p\rangle \simeq0.1$ throughout the nonlinear stage.

Fig.~\ref{fig:fft} (panel $(a)$) shows the isotropic power spectra of the poloidal ($B_\mathrm{pol}\equiv\sqrt{B_x^2+B_z^2}$) and toroidal ($B_y$) magnetic field with respect to perpendicular wavenumbers $k_\perp$, averaged over $t\in[7,8]P_0$ during the nonlinear stage. Our simulation allows us to attain a decade-wide separation between MHD and kinetic scales, observable in this figure: energy is injected around the most-unstable MRI wavenumber~$k_\mathrm{MRI}$; a distinct inertial range is identified by characteristic power laws $\propto k_\perp^{-2}$ (for $B_y$) and $\propto k_\perp^{-5/3}$ or $\propto k_\perp^{-3/2}$ (for $B_\mathrm{pol}$) at $k_\perp\langle\rho_\mathrm{c}\rangle<1$. The spectral slopes in the inertial range are compatible with earlier MHD and hybrid studies \cite{kunz2016,walker2016}. We then observe a spectral break around $k_\perp\langle\rho_\mathrm{c}\rangle=1$, followed by a modification of the spectra to power laws steeper than $\propto k_\perp^{-3}$.

Panel $(b)$ of Fig.~\ref{fig:fft} shows the isotropic power spectra of compressive ($\vecV_\mathrm{compr}$) and solenoidal ($\vecV_\mathrm{sol}$) bulk-velocity fluctuations, calculated from $\vecV\equiv\bb{U}/\Gamma=\vecV_\mathrm{compr}+\vecV_\mathrm{sol}$ such that $\grad\btimes\vecV_\mathrm{compr}=\bb{0},\grad\bcdot\vecV_\mathrm{sol}=0$. We observe that the $\bb{V}_\mathrm{sol}$ spectrum possesses a spectral slope $k_\perp^{-2}$ in the MHD range; $\bb{V}_\mathrm{compr}$ follows a roughly similar scaling, although numerical noise makes it harder to identify a clear spectral slope. These spectra are fundamental for identifying the type of turbulent fluctuations that contribute to plasma heating, including differential-heating mechanisms of electron-ion plasmas in low-luminosity accretion flows \cite{kawazura2020,zhdankin2021comp,kawazura2022}. Here, we observe that solenoidal (predominantly Alfv\'enic) fluctuations strongly dominate, in terms of spectral power, at all scales (including sub-Larmor scales, where, however, numerical noise impacts the results). This is expected, since the MRI acts, in general, as a solenoidal driver for turbulence \cite{gong2020}.

Particle energy spectra $f(\gamma)\equiv\rmd N/\rmd\gamma$ are shown in Fig.~\ref{fig:dist} for $t\in[4,15]P_0$, i.e.,\ during the nonlinear stage. When SC is turned off at $t=4P_0$, the distribution lacks strong nonthermal features\footnote{In contrast with previous studies, where the initial distribution of the turbulence stage was already strongly nonthermal.}. We then observe the effect of MRI turbulence on this initial distribution: Fig.~\ref{fig:dist} (panel $(a)$) showcases continuous plasma heating, indicated by a shift of the distribution peak toward higher energies over time. Concurrently, a steep nonthermal tail $f(\gamma)\sim\gamma^{-4}$ develops and evolves toward $f(\gamma)\sim\gamma^{-2.5}$ at the end of the run. At that point, particles have accumulated around the highest possible energy determined by $\rho_\mathrm{c}\sim L_z$ as also observed in previous MRI simulations \cite{bacchini2022mri}. The power-law tail is less steep than in externally driven-turbulence simulations of similar plasma beta~\cite{zhdankin2017}. For comparison, in panel $(b)$ we show spectra for an equivalent simulation (same initial parameters except for $\Delta x\simeq c/\omp$), where SC remains active throughout the nonlinear stage. Here, $f(\gamma)$ settles onto a low-energy Maxwellian state that is maintained indefinitely, resulting from a balance between energy gain from turbulence and energy loss from~SC.
This is consistent with previous externally driven turbulence simulations of strongly radiative pair plasmas \cite{zhdankin2020, groselj2023}. Despite the lack of NTPA, the radiative and nonradiative simulations proceed almost indistinguishably, in terms of magnetic-field amplification, average pressure anisotropy, and turbulent spectra (not shown), at least over the simulated time scale $t\leq 15P_0$.

\prlsection{Discussion}
We have presented the largest (to date) 3D PIC-SB simulations of collisionless pair-plasma MRI turbulence, with and without radiation-reaction forces. Our unprecedentedly large runs are critical for determining the \textit{mesoscale} steady-state behavior of the 3D MRI from first principles. We found that magnetic-field amplification and overall MRI dynamics, from the initial linear instability to the quasi-stationary nonlinear turbulence, broadly align with MHD and hybrid expectations: a saturated $\beta\sim10$ is maintained throughout the nonlinear stage, with plasma slowly heating due to cascading turbulence from MHD down to kinetic scales. However, our fully kinetic runs also add novel, fundamental information to the picture.

For the first time, we obtain an MHD-like, fully kinetic ordering of stresses contributing to AMT in collisionless MRI turbulence. Braginskii-MHD simulations \cite{sharma2006,kempski2019} have shown that substantial AMT can occur in anisotropic-pressure fluid models. Those simulations necessarily employ ad-hoc closures from kinetic plasma theory to limit pressure-anisotropy growth. Enhanced AMT was also observed in hybrid (fluid electrons/kinetic ions) SB simulations \cite{kunz2016}. These earlier models provided an expectation for the hierarchy of the different stresses contributing to AMT, showing that $A_{xy}\leq M_{xy}$. All previous 2D/3D PIC-MRI studies \cite{riquelme2012,hoshino2013,hoshino2015,inchingolo2018,bacchini2022mri} showed AMT enhancement, but often with a reversed ordering $M_{xy}\leq A_{xy}$ especially during the late nonlinear stage, where $M_{xy}\ll A_{xy}$ was typically observed. This was likely due to limited separation between macro- and microscopic scales. The present study employs a scale separation $\omrat$ twice as large as  employed among all previous 3D works \cite{bacchini2022mri}; this proved sufficient to recover the expected fluid trend $A_{xy}\lesssim M_{xy}$.
We also showed that pressure anisotropy, regulating $A_{xy}$, remains well-controlled by fast mirror and firehose instabilities, which are naturally included in our first-principles model.
Our results could be used as an input, e.g., for (global/local) Braginskii-MHD simulations of accretion, replacing ad-hoc mirror/firehose thresholds with PIC-derived functions of local plasma conditions.

Exploiting our large scale separation, we analyzed the MRI-turbulent magnetic and velocity-fluctuation spectra from macroscopic to microscopic scales, over two decades in $k_\perp$. Magnetic spectra show a clear steepening of spectral slopes, from $\sim -2$ (for $B_y$) and $\sim -5/3$ or $\sim-3/2$ (for $B_\mathrm{pol}$) at large scales, aligning with MHD expectations \cite{walker2016}, to slopes $<-3$ at kinetic scales. It is presently difficult to compare our results with existing analytic/simulation expectations, due to our specific plasma conditions and injection time scales characteristic of the MRI dynamics. Standard pair-plasma-turbulence simulations have shown a steepening of magnetic spectra at kinetic scales broadly aligning with ours \cite{zhdankin2017}; in the future, we will conduct thorough comparisons between these results. Velocity-fluctuation spectra display a spectral slope $\sim -2$ in the inertial range, steeper than the ideal-MHD expectation of $-5/3$. This could be attributed to the effect of anisotropic viscous stresses damping velocity fluctuations, as observed in Braginskii-MHD simulations \cite{kempski2019}.
Earlier reduced-MHD MRI-turbulence estimates showed a dominance of slow (i.e.\ compressive) over Alfv\'enic fluctuations \cite{kawazura2022}, which may influence particle energization at small scales.
Applying the same diagnostic in our simulations would be interesting but less than straightforward, since a clear way to distinguish between these modes in PIC runs is not available in literature. In addition, those previous MHD studies considered a substantially different setup with a strong, unidirectional toroidal field throughout the disk. This is not the case for our runs, where a strong mean $B_y$ is absent. A thorough comparison between full PIC and reduced-MHD results is beyond the scope of this work but certainly worth pursuing in future work.

Finally, the main result of this study is the first definitive demonstration that MRI turbulence can efficiently accelerate particles to nonthermal energies. Our quantification is, for the first time, \textit{agnostic of initial nonthermal particle injection}, i.e.\ we completely separated NTPA due to steady-state MRI turbulence from ``spurious'' NTPA due to initial MRI stages, where nonthermal particles are preaccelerated by magnetic reconnection associated with the collapse of transient macroscopic channel flows. We suppressed this effect by including SC at preturbulence stages, and turning SC off upon reaching the nonlinear stage. We then followed the evolution of particle energy spectra from a ``clean'', near-Maxwellian state, measuring significant NTPA creating high-energy power-law tails. This also marks the first time that radiation reaction is included self-consistently in kinetic SB-MRI simulations. For demonstration, we also conducted a full simulation where SC is kept active at all times; there, a stationary thermal distribution arises from the balance between MRI energy injection and SC backreaction, as expected.
Our addition of SC is fundamental for future, more realistic electron-ion semirelativistic MRI scenarios [E. Gorbunov et al., in prep.], where electron SC must be included in simulations to correctly interpret SMBH observations \cite{EHT2019e,EHT2022e}.

\begin{acknowledgments}
F.B.\ would like to thank Benjamin Crinquand for discussions and guidance on the implementation of radiation-reaction forces. F.B.\ also thanks Lorenzo Sironi, Giuseppe Arr\`o, Mario Riquelme, and Astor Sandoval for useful discussions throughout the development of this work.
F.B.\ acknowledges support from the FED-tWIN programme (profile Prf-2020-004, project ``ENERGY'') issued by BELSPO, and from the FWO Junior Research Project G020224N granted by the Research Foundation -- Flanders (FWO).
The authors acknowledge support from a NASA ATP grant 80NSSC20K0545.
Additional support was provided by National Science Foundation grants AST 1806084 and AST 1903335 to the University of Colorado.
Support for L.A.\ was provided by the Institute for Advanced Study.
Support for V.Z.\ at the Flatiron Institute is provided by the Simons Foundation.
This work was performed in part at Aspen Center for Physics, which is supported by National Science Foundation grant PHY-1607611.

This work made use of substantial computational resources provided by several entities. We acknowledge support by the VSC (Flemish Supercomputer Center), funded by the Research Foundation -- Flanders (FWO) and the Flemish Government -- department EWI.
An award of computer time was provided by the Innovative and Novel Computational Impact on Theory and Experiment (INCITE) program, using resources (namely Theta) of the Argonne Leadership Computing Facility, which is a DOE Office of Science User Facility supported under contract DE-AC02-06CH11357.
Additional computer time was provided by the Extreme Science and Engineering Discovery Environment (XSEDE), which is supported by National Science Foundation grant number ACI-1548562 \citep{towns2014}, and by the Frontera computing project \citep{stanzione2020} at the Texas Advanced Computing Center (TACC, \href{http://www.tacc.utexas.edu}{www.tacc.utexas.edu}). We acknowledge TACC for providing HPC resources on both Stampede2 and Frontera.
Additional resources (namely Pleiades) supporting this work were provided by the NASA High-End Computing (HEC) Program through the NASA Advanced Supercomputing (NAS) Division at Ames Research Center.
\end{acknowledgments}

\input{paper_MRI_PRL.bbl}

\end{document}

%% file: paper_MRI_PRL.bbl
%